\documentclass[epj]{svjour}
\usepackage{graphicx,amsfonts,amsmath}
\usepackage{bbm,color}
\newcommand{\smb}{{\scriptscriptstyle B}}
\newcommand{\sma}{{\scriptscriptstyle A}}
\def\id{\mathbbm I}
\begin{document}
\title{The data aggregation problem in quantum hypothesis testing}
\author{Simone Cialdi \inst{1,2}, Matteo G. A. Paris \inst{1,3}}
\institute{
\inst{1} Dipartimento di Fisica dell'Universit\`a degli 
Studi di Milano, 20133 Milano, Italia \\ 
\inst{2} INFN, Sezione di Milano, I-20133 Milano, Italy \\ 
\inst{3} CNISM, UdR Milano Statale, I-20133 Milano, Italy.}
\date{\today}
\abstract{
We discuss the implications of quantum-classical Yule-Simpson effect
for quantum hypothesis testing in the presence of noise, and provide 
an experimental demonstration of its occurrence in the problem of 
discriminating which polarization quantum measurement has been actually 
performed by a detector box designed to measure linear polarization of 
single-photon states along a fixed but unknown direction.}
\PACS{{03.67.-a}, {03.65.Wj}, {42.50.Ex}, {02.50.-r}}
\maketitle
\section{Introduction}
\label{s:ys}
The Yule-Simpson effect \cite{Sim51,Bly72,Bic75} in statistics 
occurs when the correlations observed within different samples 
are reversed when the sampled are combined together. 
Although no actual mathematical paradox is involved, the 
Yule-Simpson effect has an impact on statistical inference, 
since the aggregated data and the partitioned ones may 
suggest opposite conclusions.
Two forms of the Yule-Simpson effect in quantum measurements has been
recently introduced in \cite{qs} and their occurrence in qubit systems 
have been experimental verified \cite{qse}. The possible connections 
of the effect with high order Bell-Tsirelson inequalities have been 
also explored \cite{qsb}.
\par
In this paper we discuss the implications of quantum-classical 
Yule-Simpson effect for quantum hypothesis testing in the presence 
of noise. In particular, we demonstrate its occurrence in the problem 
of discriminating which polarization quantum measurements has been 
actually performed by a given {\em detector box}, 
designed to measure linear polarization of 
single-photon states along one of two possible directions.
\par
Suppose that you are given a box, which may implement two possible 
dichotomic measurements $A=\{\Pi_\sma, \id-\Pi_\sma\}$ and $B=\{\Pi_\smb, 
\id-\Pi_\smb\}$ on a given system, and you have to infer 
which  measurement has been performed on the basis of the results of the 
measurement. To this aim, you may  {\em probe} the measuring box 
$M$ times by suitably prepared states of the system. In our scheme
the box is performing (linear) polarization 
measurements along a given direction, or along a slightly 
tilted one. Let us denote by $\theta$ the possible tilting angle. 
The two measurements are thus described by the operator measures
$\Pi_\sma=|0\rangle\langle0|$, $|0\rangle$ describing vertical
polarization in the given direction,
or $\Pi_\smb=|0\rangle_{\theta}{}_{\theta}\langle0|$, where
$|0\rangle_{\theta}=\cos\theta |0\rangle + \sin\theta |1\rangle$
\cite{parqm}.
\par
In order to discriminate which measurement has been actually
performed, one sends some probe signal and take a decision on the
basis of the measurement results. As for example, we may send
photons with a definite polarization state, e.g. 
$\varrho_0=|0\rangle\langle 0|$, 
corresponding to linear vertical polarization along the given 
direction. In this case the detector always returns the "$0$" outcome if
the box is performing $A$ measurement, while some fraction of 
"$1$" is expected in case of the $B$ measurement. More precisely, 
the probabilities of obtaining the outcome "$0$" with the two
measurements are given by 
\begin{align}
p_1&=\langle 0|\Pi_\sma |0\rangle =1 \notag \\ 
q_1&=\langle 0|\Pi_\smb |0\rangle =|\langle 0|0\rangle_\theta|^2=\frac12 (1+\cos2\theta) 
\label{p1s}
\end{align}
Let us now admit that some external perturbation may introduce some
noise in the preparation stage of the probe signal. In particular, 
we assume that if the noise is present then the probe is prepared 
in a mixture of states having linear vertical polarization along 
a random direction, tilted by 
small angle $\alpha$ from the given axis. In order to make minimal
assumptions on the nature of the perturbation, we take the angles
$\alpha$ distributed according to a Gaussian with zero mean.
In this case the polarization state of the probing photons is 
described by the density operator
\begin{equation}
\varrho_\Delta \equiv {\cal D}_\Delta (\varrho_0)
= \int\!\! d\alpha\, \frac{e^{-\frac{\alpha^2}{2 \Delta^2}}}{\sqrt{2 \pi
\Delta^2}}\, |0\rangle_\alpha{}_\alpha\langle 0| \,,
\label{gd}
\end{equation}
where $\Delta\ll 2 \pi$, such that the integral may be safely evaluated over
the entire real axis. The probabilities of getting the "$0$" outcome
for the two measurements are now given by 
\begin{align}
p_2&=\langle 0 |\varrho_\Delta | 0\rangle = \frac12 (1+\delta) \notag \\ 
q_2&={}_\theta\langle 0 |\varrho_\Delta | 0\rangle_\theta = \frac12
(1+\delta \cos2\theta) 
\label{p2s}\,,
\end{align}
where $\delta=\exp(-2\Delta^2)$ represents the smearing effects
of the preparation noise.
\par
At first sight, the presence of preparation noise is not changing the
picture. Indeed, we have that $p_2$ is larger that $q_2$, such that
one still expects a larger number of "$0$" outcomes when the box is
performing the $A$ measurement. 
\par
On the other hand, and this is the {\em data aggregation problem}
that we mention in the title of the paper, if we do not know how many
times the perturbation in the preparation state occurred, it could happen that
the overall probability of the event "$0$" is larger for the $B$
measurement than for the $A$ measurement, i.e. we may expect more
"$0$" by measuring polarization along the tilted direction than with the
original one.  In order to 
understand how this may happen, let us denote by $\gamma=M_0/M$ the fraction of 
runs where the box is probed by the state 
$\varrho_0$. 
The overall  density operator describing the polarization state of the probing 
photon is given by 
\begin{equation}
\varrho_\gamma \equiv \Phi_{\gamma\Delta} (\varrho_0)
= \gamma\, \varrho_0 + (1-\gamma)\, \varrho_\Delta
\label{vg}\,
\end{equation}
which may be seen as the output state from an overall two-parameter
noisy channel described by the map 
\begin{equation}
\Phi_{\gamma\Delta} = \gamma {\cal I} + (1-\gamma) {\cal D}_\Delta\,,
\label{vg1}
\end{equation}
being ${\cal I}$ the identity channel and 
${\cal D}_\Delta$ the phase-diffusion one, introduced in Eq. (\ref{gd}).
\par
The probabilities of the "$0$" outcome for the two measurements
is given by
\begin{align}
p&=\langle 0 |\varrho_\gamma | 0\rangle = \gamma\, p_1 +(1-\gamma)\, p_2 \notag \\ 
q&={}_\theta\langle 0 |\varrho_\gamma | 0\rangle_\theta = \gamma\, q_1 +
(1-\gamma)\, q_2
\label{ps}\,.
\end{align}
The data aggregation problem consists in the fact that there exist
frequencies $\gamma_1$ and $\gamma_2$ such that 
$\gamma_2 q_1+(1-\gamma_2) q_2 > \gamma_1 p_1 + (1-\gamma_1) p_2$
despite the fact that $p_1>q_1$ and $p_2>q_2$. This happens
if $$\gamma_2 > \frac{p_1-p_2}{q_1-q_2}\, \gamma_1 + \frac{p_2-q_2}{q_1-q_2}\,,$$ 
i.e.
$$
\gamma_2 > \frac{\gamma_1}{\cos2\theta} + \frac{\delta}{1-\delta} 
\frac{1-\cos2\theta}{\cos2\theta}\,.
$$
Remarkably, the above relation may be satisfied by some pairs of
frequencies $\gamma_1$ and $\gamma_2$ whenever $\delta< 2 \cos 2\theta$.
For fixed frequencies the effect takes place if the preparation noise is 
larger than a threshold, corresponding to
$$
\delta < \frac{\gamma_1 - \gamma_2 \cos 2\theta}{\gamma_1-1
-(\gamma_2-1) \cos 2\theta} \stackrel{\theta\ll 1}{\simeq} 1 -
\frac{2\theta^2}{\gamma_2 - \gamma_1}
$$
\par
Summarizing, we probe the detector box by a pair of possible 
preparations, described by the density operators $\varrho_0$
and $\varrho_\Delta$, corresponding to negligible 
noise acting on the probe ($\varrho_0$) or to the presence of 
non-negligible noise described by Gaussian mixing ($\varrho_\Delta$). 
After the measurement, 
we aim to infer which polarization has been
actually measured on the basis of the number of, say, $"0"$ outcomes
recorded after $M=M_0+M_\Delta$ repeated measurements, where $M_j$ 
is the number of runs where the system was prepared in the state
$\varrho_j$, $j=0,\Delta$. 
If we know which preparation $\varrho_j$ has been used in 
each run, i.e. we know when the noise is present, then we are able 
to make a definite inference, say $A$ measurement if $p_j>q_j$,  
independently on the number of runs.
On the other hand, if we ignore the information about which preparation 
has been sent to the box in each run, i.e. we aggregate data because we
do not know whether the noise was present or not, then 
we may reach the opposite conclusion, depending on the relative weight
$M_0/M_\Delta$ of the samples.  This is a manifestation of the
quantum-classical Yule-Simpson effect, which may easily occur when
discriminating measurement apparatuses in the presence of 
noisy channels described by maps of the form (\ref{vg1}).
Overall, there is no
mathematical paradox: still the aggregated data and the 
partitioned ones may, in fact, suggest opposite conclusions.
The effect is referred to as {\em quantum-classical} YS effect
since it occurs in quantum measurements due to classical uncertainty
in the preparation of the probe signals, i.e. to the presence of 
mixed probes. An analogue {\em quantum-quantum} YS effect may indeed
occur with superpositions \cite{qs}.
\par
In the next Sections we describe and discuss an experimental 
scheme where the above effect takes place. 
\section{Experimental apparatus}
\begin{figure}[h!]
\includegraphics[width=0.4\columnwidth]{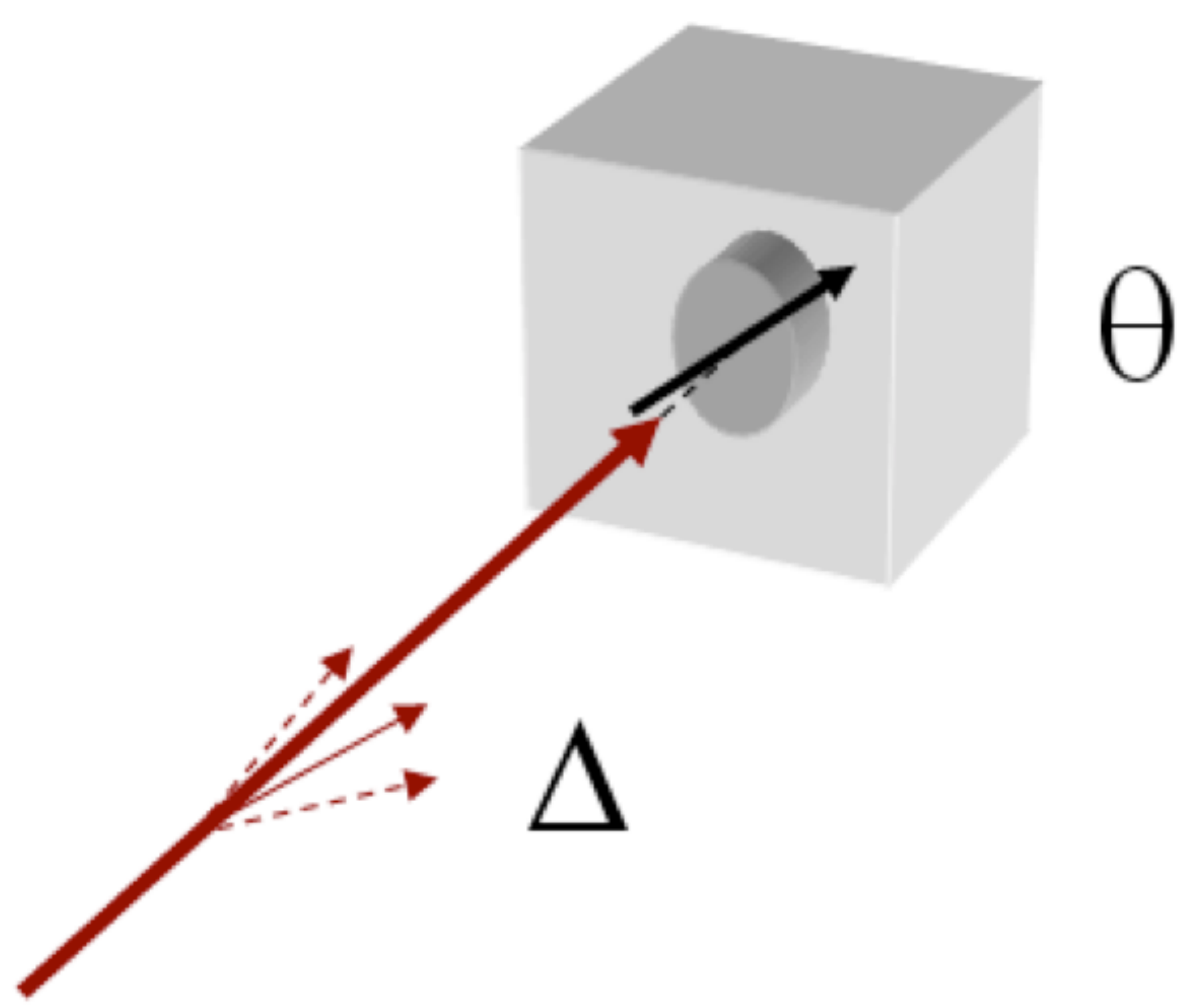}
\includegraphics[width=0.58\columnwidth]{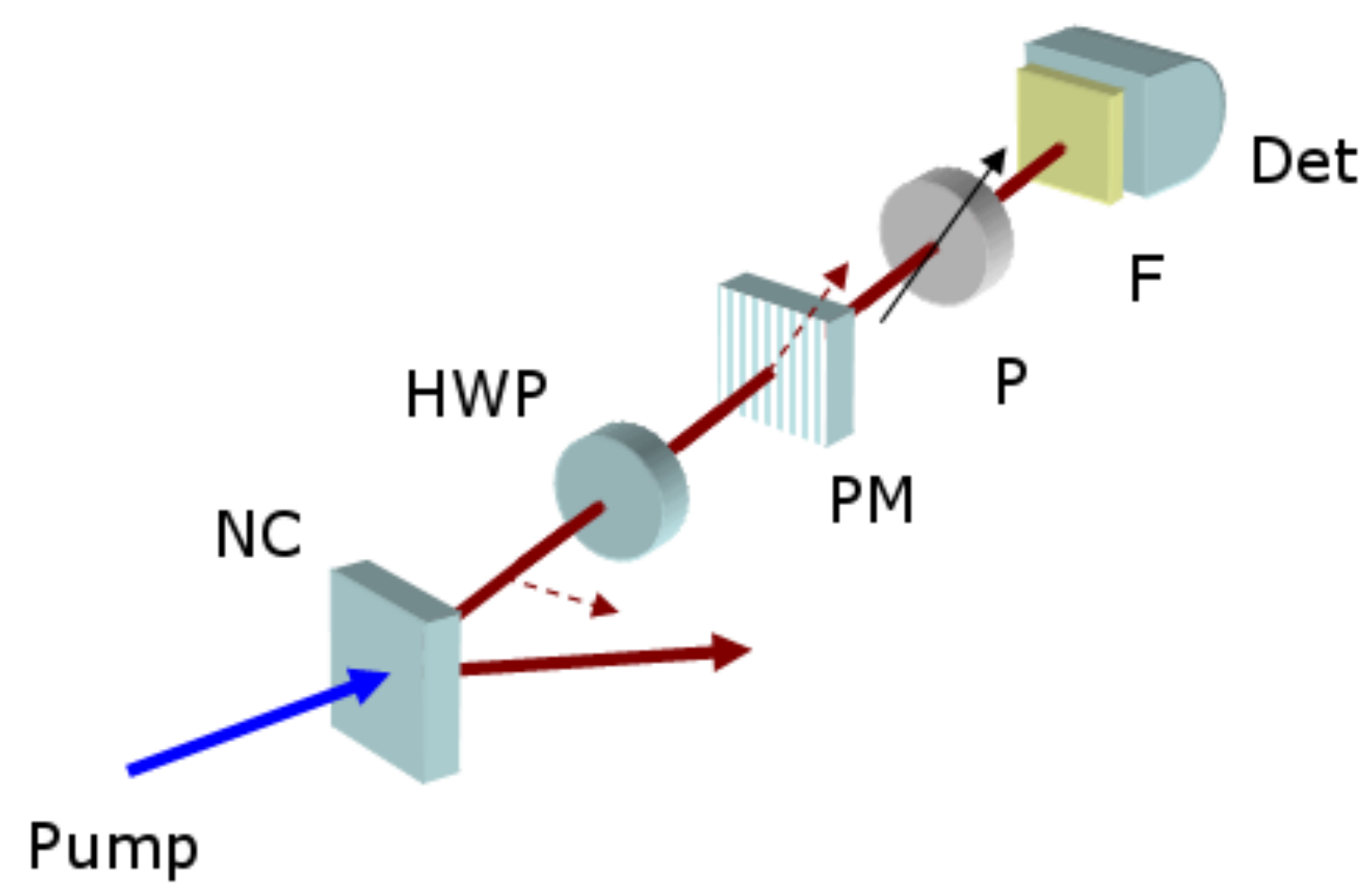}
\caption{
(Color online) (Left): Basic blocks of the experiment. A light beam  
which {\em may} be subject to Gaussian polarization diffusion ($\Delta$ is 
the standard deviation)
enters in a detector box which contains a linear 
polarization analyzer set at an angle $0$ {\em or} $\theta$ with respect 
to a reference axis. (Right): Schematic 
diagram of the experimental apparatus. A $\beta$-barium borate crystal 
(NC, length $3$ mm), pumped by a linearly polarized cw $405$ nm diode 
laser, is the source of horizontally polarized photon pairs via parametric 
down-conversion.  Then the polarization is set at $45^\circ$ by an 
half-wave-plate (HWP). The ideal scheme is simulated introducing a proper 
phase shift by a phase modulator (PM) and a polarizer (P) set a
$45^\circ$. (F) is a long-pass filter (cut-on
wavelength = $780$ nm) and (Det) is a single photon detector.}
\label{f:sch}
\end{figure}

The logical scheme of the experiment, corresponding to 
the situation described in Section \ref{s:ys}, is shown in the
left panel of Fig. \ref{f:sch}, whereas the experimental setup is shown 
in the right panel of the same figure. We work with photon polarization
since this is a degree of freedom which may be reliably controlled. 
In turn, it has been already shown that the noise model introduced in 
the previous section may be reliably implemented \cite{nm1,nm2}. 
\par
A linearly polarized cw $405$ nm diode laser (Newport LQC405-40P) pumps a 
$\beta$-barium borate crystal (NC, length $3$ mm) cut for type-I down 
conversion with the optical axes aligned in the horizontal plane. The 
non-linear crystal is used  as a source of horizontally polarized photon 
pairs via parametric down conversion. We use an half-wave-plate (HWP) 
to set the polarization at $45^\circ$. Then, in order to obtain a scheme 
equivalent to that of the left panel of Fig. \ref{f:sch}, we use a phase 
modulator and a polarizer set at $45^\circ$ (see below). Finally, we have 
a long-pass filter (cut-on wavelength = $780$ nm) to reduce the background 
and an home-made single photon detector (Det).
With the phase modulator it is possible to introduce an arbitrary phase
shift $\phi$ between the horizontal (H) and the vertical (V) polarization. 
After the polarizer set a $45^\circ$ the probability to see a
photon is thus $\frac12(1+\cos\phi)$. 
The acquisition consist of $200$ iterations. For each iteration we acquire
$4$ counts, each within a temporal window of $1$ second: $N_{1p}$ are the
counts obtained for $\phi=0$, $N_{1q}$ are for the setting $\phi=2\theta$, 
$N_{2p}$ corresponds to $\phi=-2\alpha$, and $N_{2q}$ to $\phi=2(\theta-\alpha)$, 
where $\alpha$ is randomly
sampled from a Gaussian distribution of zero mean and 
variance $\Delta$. Since, according to Eq.(\ref{p1s}), for $\phi=0$ we have 
$p_1=1$, then $N_{1p}$ is used as a normalization to estimate the other 
probabilities as follows: 
$$q_1=N_{1q}/N_{1p} 
\quad p_2=N_{2p}/N_{1p} \quad q_2=N_{2q}/N_{1p}\,.$$
After the acquisition of the four counts, we emulate the lack of
knowledge about the preparation of the probe by mixing the $N_p$ 
and the $N_q$ data according to a pair of dichotomic distributions 
$(\gamma_1, 1-\gamma_1)$ and $(\gamma_2, 1-\gamma_2)$. We thus
obtain {\em online} the ratios $q_1/p_1$ and $q_2/p_2$, as well as $q/p$, together
with their corresponding uncertainties.
\section{Results}
\begin{figure}[h!]
\includegraphics[width=0.48\columnwidth]{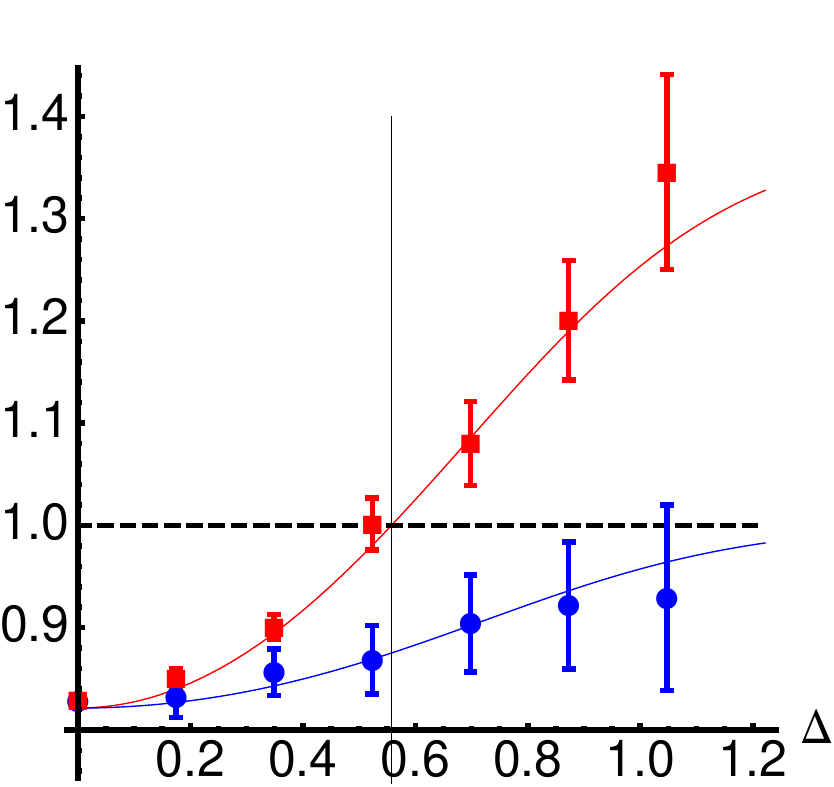}
\includegraphics[width=0.48\columnwidth]{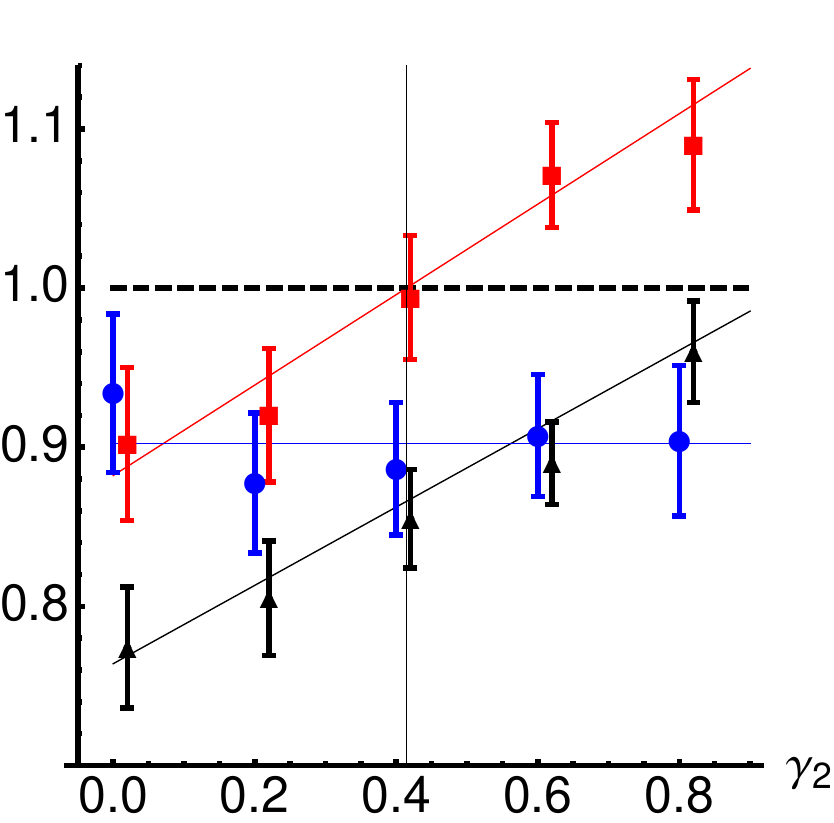}
\caption{
(Color online) (Left): the ratios $q_2/p_2$ (blue circles) and $q/p$
(red squares) as a function of the
preparation noise parameter $\Delta$ for fixed values of the frequencies
$\gamma_1=0.1$, $\gamma_2=0.8$ and for the alternative measurement taken
at $\theta=\frac{5}{36}\pi$ rad. Solid lines denotes the theoretical
predictions of Eqs. (\ref{p2s}) and (\ref{ps}). The vertical line 
denotes the noise threshold for the occurrence of the YS effect at the
given values of $\gamma_1$, $\gamma_2$, and $\theta$, i.e. 
$\Delta_{\text{th}}\simeq 0.558$ rad.
(Right): 
the ratios $q_2/p_2$ (blue circles) and $q/p$
(red squares for $\gamma_1=0.05$ and black triangles for $\gamma_1=0.4$) 
as a function of probability $\gamma_2$ for a fixed value
of the preparation noise $\Delta=\frac{2}{9}\pi$ rad and for 
the alternative measurement taken
at $\theta=\frac{5}{36}\pi$ rad.
Solid lines denotes the theoretical
predictions of Eqs. (\ref{p2s}) and (\ref{ps}). The vertical line 
denotes the threshold for the occurrence of the YS effect 
for $\gamma_1=0.05$ and at the given values of $\theta$ and 
$\Delta$, i.e. $\gamma_2=0.414$ (no YS effect for $\gamma_1=0.4$).
Notice that the $q/p$ data in the right panel have been slightly shifted
to the right for clarity but they have been collected for the same
values of $\gamma_2$ as the $q_2/p_2$ ones. 
}
\label{f:f2}
\end{figure}
\par
Experimental results are summarized in Fig. \ref{f:f2}. In the left
panel we show the ratios $q_2/p_2$ (blues circles) and $q/p$ (red
squares) as a function of the preparation noise parameter $\Delta$ 
for fixed values of the frequencies
$\gamma_1=0.1$, $\gamma_2=0.8$ and for the alternative measurement taken
at $\theta=25^\circ = \frac{5}{36}\pi$ rad (the ratio $q_1/p_1$ is smaller than unit by
construction). Data are in excellent agreement with 
the theoretical predictions of Eqs. (\ref{p2s}) and (\ref{ps}) (solid
lines) and confirm the occurrence of the YS
effect in quantum hypothesis testing in the presence of noise. For our 
choice of $\gamma_1$, $\gamma_2$ and $\theta$ the noise threshold for the 
YS effect was
$\Delta>\Delta_{\text{th}}\simeq 0.558$ rad.
\par
In the right panel we show the ratios $q_2/p_2$ (blue circles) and $q/p$
(red squares for $\gamma_1=0.05$ and black triangles for $\gamma_1=0.4$) 
as a function of probability $\gamma_2$ for a fixed value
of the preparation noise $\Delta=\frac{2}{9}\pi$ rad and for 
the alternative measurement taken at $\theta=\frac{5}{36}\pi$ rad.
Data are in excellent agreement with the theoretical predictions of 
Eqs. (\ref{p2s}) and (\ref{ps}) (solid lines), confirming that $q_2/p_2$
is independent on the choice of the probabilities $\gamma_1$ and
$\gamma_2$, and showing that YS effect may occur for increasing
$\gamma_2$.
The vertical line 
denotes the threshold for the occurrence of the YS effect 
for $\gamma_1=0.05$ and at the given values of $\theta$ and 
$\Delta$, i.e. $\gamma_2=0.414$, whereas, as expected, no YS effect occurs 
for $\gamma_1=0.4$.
\par
\section{Conclusions}
In this paper, we have discussed the implications of quantum-classical 
Yule-Simpson effect for quantum hypothesis testing and demonstrated
its occurrence in the problem of discriminating which polarization 
quantum measurements has been actually performed by a given box, with 
the two possible detectors designed to measure linear 
polarization of single-photon states 
along slightly different directions.
If noise affects the preparation stage, one is actually 
probing the box with two different kinds of signals, the unperturbed 
one and its noisy version. Since one
usually ignores which preparation actually arrived at the detector in each 
run, data from the two preparations are {\em aggregated} and one may reach 
opposite inference, depending on the noise occurrence rate.  This is a 
plain manifestation of the
quantum-classical Yule-Simpson effect, which may easily occur when
discriminating measurement apparatuses in the presence of noise.
Overall, there is no mathematical paradox: still the effect is
puzzling for what concerns statistical
inference, since the aggregated data and the 
partitioned ones may, in fact, suggest opposite conclusions.
\section*{Acknowledgment}
This work has been supported by the MIUR project FIRB-LiCHIS-RBFR10YQ3H. 

\end{document}